# Efficient and Robust Propensity-Score-Based Methods for Population Inference using Epidemiologic Cohorts

Lingxiao Wang,[1,2] Barry I. Graubard,[2] Hormuzd A. Katki,[2] and Yan Li[1,*]

[1]The Joint Program in Survey Methodology, University of Maryland, College Park, U.S.A.

[2]National Cancer Institute, Division of Cancer Epidemiology & Genetics, Biostatistics Branch, U.S.A.

**YL*: 1218 Lefrak Hall, 7521 Preinkert Dr, College Park, MD 20742 yli6@umd.edu

**SUMMARY (183 words)**

Most epidemiologic cohorts are composed of volunteers who do not represent the general population. To enable population inference from cohorts, we and others have proposed utilizing probability survey samples as external references to develop a propensity score (PS) for membership in the cohort versus survey. Herein we develop a unified framework for PS-based weighting (such as inverse PS weighting (IPSW)) and matching methods (such as kernel-weighting (KW) method). We identify a fundamental Strong Exchangeability Assumption (SEA) underlying existing PS-based matching methods whose failure invalidates inference even if the PS-model is correctly specified. We relax the SEA to a Weak Exchangeability Assumption (WEA) for the matching method. Also, we propose IPSW.S and KW.S methods that reduce the variance of PS-based estimators by scaling the survey weights used in the PS estimation. We prove consistency of the IPSW.S and KW.S estimators of population means and prevalences under WEA, and provide asymptotic variances and consistent variance estimators. In simulations, the KW.S and IPSW.S estimators had smallest MSE. In our data example, the original KW estimates had large bias, whereas the KW.S estimates had the smallest MSE.

Keywords: Nonprobability cohorts, finite population inference, propensity score weighting, weight scaling, Taylor series linearization variance.

# 1 INTRODUCTION

Assembling epidemiologic cohorts using probability sampling substantially increases costs, especially if biospecimens are required (LaVange, Koch, and Schwartz, 2001; Duncan, 2008). Thus, many cohort studies assemble samples using volunteer-based recruitment, but they generally cannot represent the general population. For example, the UK Biobank has half the mortality rate of the UK population (Collins, 2012). However, the recent survey sampling literature is burgeoning with propensity-score (PS)-based methods that use probability-based survey samples as external references to improve population inferences for non-probability samples (Lee and Valliant, 2009; Valliant and Dever, 2011; Elliot and Valliant, 2017; Chen, Li, and Wu, 2019). Intuitively, these methods create pseudo-weights for the cohort participants so that the underrepresented are upweighted, and the overrepresented are down weighted.

One popular type of PS-based method is PS-based weighting methods that use PS to estimate participation rates of the cohort units, for example, the inverse PS weighting (IPSW) (Valliant and Dever, 2011). A second approach is PS-based matching methods that use PS to measure similarity of the cohort and survey sample units, such as PS adjustment by subclassification (PSAS) (Lee and Valliant, 2009), Rivers' matching method, (Rivers, 2007), and Kernel Weighting (KW) (Wang et al., 2020). Compared to PS-based weighting methods, PS-based matching methods can be more robust to propensity model mis-specification because the PS is used to measure the similarity of the cohort with the survey sample units, in terms of the distribution of the covariates that are considered in the propensity model (Wang et al., 2020). In PS-based matching methods, the propensity model is fitted to the combined (cohort vs. unweighted survey) sample to avoid inefficiency in estimation due to using the survey weights (Lee and Valliant, 2009; Brick, 2015; Rivers, 2007; Wang et al., 2020). The KW method, which uses kernel smoothing, ensures consistent pseudo-weighted estimates under regularity conditions (Wang et al., 2020). Empirical results generally show that the KW estimation of finite population means generally have smaller mean squared error (MSE) compared to IPSW or PSAS estimation. In this paper, we focus on IPSW and KW as examples of the PS-based weighting and matching methods.

In this paper, we demonstrate that all PS-based matching methods that fit the propensity model to the combined (cohort vs. unweighted survey) sample require a hidden, but critical, strong exchangeability assumption (SEA) for estimating finite population means. The SEA states that the expectation of the outcome variable given the estimated PS is the same in the cohort, the survey, and the finite population. We prove that, without the SEA, current PS-based matching estimates are not consistent, even when the propensity model is correctly specified. We establish a unified framework for both PS-based weighting and matching methods. For the matching methods, the matching scores are defined by the (functions of) PS's estimated from the propensity models fitted to the combined (cohort vs. **weighted** survey) sample. We relax the SEA to a Weak Exchangeability Assumption (WEA) that is more realistic for data-analysis.

However, fitting the propensity model to the weighted sample (when compared to the unweighted sample) increases variability in PS estimation that can greatly increase the variance of pseudo-weighted estimation. To recover efficiency, we propose **scaling** survey weights by their mean in PS estimation, which is motivated by the method of scaling weights in population-based case-control studies where the sample weights are highly variable among the cases and the controls (Scott and Wild, 1986; Li, Graubard, and DiGaetano, 2011; Landsman and Graubard, 2012). We demonstrate that this simple scaling greatly reduces variance while retaining the consistency of the PS-based pseudo-weighted estimators.

Finally, we derive Taylor Linearization variances (i.e. analytical variance) for pseudo-weighted estimators of finite population means that take the estimation of PS into account. We apply our methods to an example where we use the naïve (unweighted) US National Health and Nutrition Examination (NHANES) III as the "cohort" and the sample weighted 1997 US National Health Interview Survey (NHIS) as the reference survey. The use of naïve NHANES allows for clear assessment of the bias reduction obtained by the proposed PS-based methods without being obscured by differences in the population coverage and measurement errors between the cohort and the reference survey interfering to any large extent.

## 2 BASIC SETTING

Let the target finite population ($FP$) consist of $N$ individuals indexed by $i \in \{1, \cdots, N\}$, where each individual $i$ has values for the outcome variable of interest $y_i$ and for the vector of covariates $x_i$. We focus on estimating the $FP$ mean of $y$, i.e., $\mu = N^{-1} \sum_{i \in FP} y_i$. Let $s_c \subset FP$ denote a cohort with $n_c$ individuals. We define a random indicator variable $\delta_i^{(c)}$ ($= 1$ if $i \in s_c$; 0 otherwise) that specifies which individuals in $FP$ participate in $s_c$. Note that $FP$ and $s_c$ are also used to denote sets of indices for the target finite population and the cohort, respectively. The underlying cohort participation rate for each $i \in s_c$ is defined by

$$\pi_i^{(c)} \equiv P(i \in s_c \mid FP) = E_c\left(\delta_i^{(c)} \mid FP\right),$$

where the expectation $E_c$ is with respect to the unknown random cohort sample participation process from $FP$. The corresponding cohort implicit sample weight is $w_i = 1/\pi_i^{(c)}$ for $i \in s_c$. We require the following standard assumptions for cohort participation:

**A1**. The cohort participation indicator $\delta^{(c)}$ is independent of the outcome variable $y$ given the covariates $x$, i.e., $\Pr(\delta^{(c)} = 1 \mid y, x) = \Pr(\delta^{(c)} = 1 \mid x)$.

**A2**. All finite population units have a positive participation rate, i.e., $\pi_i^{(c)} > 0$ for $i \in FP$.

In addition, a reference survey sample $s_s$ with $n_s$ individuals is randomly selected from the $FP$. The sample inclusion indicator, inclusion probability, and the corresponding sample weights are defined by $\delta_i^{(s)}$ ($= 1$ if $i \in s_s$; 0 otherwise), $\pi_i^{(s)} = E_s\left(\delta_i^{(s)} \mid FP\right)$, and $d_i = 1/\pi_i^{(s)}$, respectively, where $E_s$ is the expectation with respect to the survey sample selection and $s_s$ also denotes the subset of indices for individuals in the survey sample from the $FP$. In practice, we assume the inclusion probabilities and sample weights are adjusted for nonresponse and calibrated for $FP$ quantities. It is assumed that the survey sampling is also independent of the outcome variable $y$ given the covariates $x$, i.e., $\Pr\left(\delta^{(s)} = 1 \mid y, x\right) = \Pr\left(\delta^{(s)} = 1 \mid x\right)$.

## 3 STRONG EXCHANGEABILITY ASSUMPTION FOR MATCHING METHODS

The PS-based matching methods use a matching score (a function of PS) to measure the similarity of the cohort and survey units in terms of the covariate distributions. Hence, they do not require the matching scores accurately estimate the participation rates $\pi_i^{(c)}$ for the cohort units. To avoid low efficiency of pseudo-weighted estimates due to high variability of the estimated pseudo-weights, the existing PS-based matching methods, including PS adjustments by the subclassification (PSAS) method (Lee and Valliant, 2009), the Rivers matching method (Rivers, 2007), and the kernel weighting (KW) method (Wang et. al., 2020), use the propensity of participating in the cohort ($s_c$) versus being selected in the survey sample ($s_s$) as the matching score, defined by,

$$\tilde{p}_i = P\{i \in s_c \mid i \in s_c \cup^* s_s\} = \frac{P\{i \in s_c \mid FP\}}{P\{i \in s_c \cup^* s_s \mid FP\}} = \frac{\pi_i^{(c)}}{\pi_i^{(c)} + \pi_i^{(s)}}. \qquad (3.1)$$

The union $s_c \cup^* s_s$ allows for duplication of individuals in both $s_c$ and $s_s$. In practice, the set of duplicates is usually small and also it is usually not possible to identify the duplicate individuals. Assume that the relationship between $\tilde{p}_i$ and $x$ follows a logistic regression model

$$\log \frac{\tilde{p}_i}{1-\tilde{p}_i} = \tilde{\boldsymbol{\beta}}^T \boldsymbol{x}_i, i \in \{s_c \cup^* s_s\}, \qquad (3.2)$$

where $\tilde{\boldsymbol{\beta}}$ is a vector of unknown regression coefficients, and can be estimated by fitting model (3.2) to the combined ($s_c$ vs. unweighted $s_s$) sample. As proved in Wang et al. (2020), using $\tilde{p}$ in the matching score requires the strong exchangeability assumption (SEA)

$$E(y \mid \tilde{p}, s_s) = E(y \mid \tilde{p}, s_c) = E(y \mid \tilde{p}, FP), \qquad (3.3)$$

for consistency of the KW estimators of finite population mean, where $E(\cdot)$ is the expectation with respect to the distribution of $y$ in $FP$. The SEA requires that two equalities hold among the three sets $s_c$, $s_s$, and $FP$. This assumption is strong and can be violated even when $\tilde{p}$ is estimated under the correct propensity model fitted to the combined ($s_c$ vs. unweighted $s_s$) sample. This is because only the first equality of the SEA automatically holds under the model (3.2) (Rosenbaum and Rubin, 1983), but the second equality may not necessarily hold, resulting in biased pseudo-weighted estimation. We use the following simple examples to illustrate the cases of unbiased mean estimation when the SEA is valid, and also the cases of biased mean estimation when the SEA is violated even if the propensity model is correctly specified.

*Simple Examples*

Suppose the covariates $x$ include two binary variables age (= 0 for young; = 1 for old), and sex (= 0 for male; = 1 for female). The distribution of $y$ depends on age and sex with the expectation $\mu_{jk} = E(y \mid \text{age} = j, \text{sex} = k)$ for $j, k = 0, 1$. We assume $\mu_{jk}$ differs by the four categories of age by sex. We assume $s_c$ and $s_s$ are two independent stratified simple random samples with the four strata defined by the age and sex groups. The implicit cohort sample weights are the same within the four strata, but different across the strata.

1. SEA Valid Case

In the SEA valid case, the values of $\tilde{p}$ are different in the four strata defined by age and sex. The SEA is satisfied because $E(y \mid \tilde{p}) = E(y \mid \text{age} = j, \text{sex} = k) = \mu_{jk}$ in $s_c$, $s_s$, and $FP$. The PS-based matching methods evenly distribute survey weights to the cohort units within each of the four matching groups (strata) defined by $\tilde{p}$ to construct pseudo-weights. As a result, the pseudo-weighted estimator of the finite population mean is unbiased.

2. SEA Invalid Case

SEA can be invalid if the value of $\tilde{p}$ cannot differentiate the four age-by-sex strata. This can happen even when the true propensity model (including both age and sex) is fitted to the combined ($s_c$ vs. unweighted $s_s$) sample. For example, if $s_c$ and $s_s$ have the same distribution of sex, the effect of sex is expected to be 0 in the fitted propensity model. As a result, the value of $\tilde{p}$ is the same within age group, regardless of sex. The second equality of SEA is violated, i.e., $E(y \mid \tilde{p}, s_c) \neq E(y \mid \tilde{p}, FP)$, since $E(y \mid \text{age}, s_c) \neq E(y \mid \text{age}, FP)$. Applying a PS-based matching method, the survey sample weights would be distributed to the cohort units only according to age, resulting in common pseudo-weights within age categories, and thus yield biased pseudo-weighted estimator of the finite population mean.

As shown by this simple example, when cohort units within subgroups have different participation rates, but the same estimated PS's, the PS-based matching methods cannot match the distribution of $y$ in the pseudo-weighted $s_c$ to that in the $FP$. This is because the second equality in SEA is invalid even if correct propensity model is fitted to the unweighted sample.

# 4 UNIFYING FRAMEWORK FOR USING PROPENSITY SCORES IN PS-BASED METHODS

We establish a unifying framework for using PS's in both PS-based weighting and matching methods. The PS-based matching methods are enhanced to relax the SEA under the framework.

Suppose we observe the covariates $x_i$ for all $i \in FP$, but we don't observe the cohort participation indicator $\delta_i^{(c)}$ for all $i \in FP$. Instead of directly modeling the cohort participation rate, $\pi_i^{(c)}$, we define $p_i = P(i \in s_c \mid i \in s_c \cup^* FP)$, where the notation $\cup^*$ represents the union of $s_c$ and $FP$ which includes the duplication of the individuals in $s_c$ that, of course, are in $FP$. Therefore, $s_c \cup^* FP$ contains $N + n_c$ individuals. Notice that $p_i \leq \frac{1}{2}$, and the equality holds only if $P(i \in s_c \mid i \in FP) = 1$. As to be shown, duplicating the units in $s_c$ is a computational device that allows us to recover an estimate of the underlying inclusion probability $\pi_i^{(c)}$ from a propensity model. According to the definition of $p_i$, we have

$$\frac{p_i}{1-p_i} = \frac{P(i \in s_c \mid i \in s_c \cup^* FP)}{P(i \in FP \mid i \in s_c \cup^* FP)} = P(i \in s_c \mid i \in FP) = \pi_i^{(c)}. \tag{4.1}$$

Assume we have a logistic regression model

$$\log\left\{\frac{p_i}{1-p_i}\right\} = \boldsymbol{\beta}^T \boldsymbol{x}_i, \text{ for } i \in s_c \cup^* FP. \tag{4.2}$$

From (4.1) and (4.2), $\pi_i^{(c)} = \exp(\boldsymbol{\beta}^T \boldsymbol{x}_i)$ allowing us to obtain the cohort participation rate via a logistic propensity model. The log-likelihood function is given by

$$l(\boldsymbol{\beta}) = \sum_{i \in s_c \cup^* FP} \{R_i \cdot \log p_i + (1 - R_i) \log(1 - p_i)\} = \sum_{i \in s_c} \log p_i + \sum_{i \in FP} \log(1 - p_i), \tag{4.3}$$

where $R_i$ indicates the membership of $s_c$ in $s_c \cup^* FP$ (i.e., $R_i = 1$ if $i \in s_c$, and $= 0$ if $i \in FP$), and the propensity score $p_i$ can be rewritten as $p_i = P(R_i = 1 \mid \boldsymbol{x}_i)$ for simplicity. Furthermore, $p_i = \text{expit}(\boldsymbol{\beta}^T \boldsymbol{x}_i)$ based on Model (4.2). Note that $\boldsymbol{\beta}$ differs from $\widetilde{\boldsymbol{\beta}}$ in Model (3.2) since the two models define the propensity differently, i.e., the probability that individual $i$ is included in $s_c$ vs. $FP$ under Model (4.2) as compared to the probability that individual $i$ is included in $s_c$ vs. $s_s$ under Model (3.2). In reality, covariates $\boldsymbol{x}_i$ can only be observed in $s_c$ and $s_s$, but not $FP$. To obtain a consistent estimator of $\boldsymbol{\beta}$, we fit the model (4.2) to the combined ($s_c$ vs. **weighted** $s_s$) where we use the sample weights $d_i$ for $i \in s_s$, in the estimation through the pseudo log-likelihood function

$$\tilde{l}(\boldsymbol{\beta}) = \sum_{i \in s_c} \log p_i + \sum_{i \in s_s} d_i \log(1 - p_i). \tag{4.4}$$

Heuristically, we are substituting the sample weighted $s_s$ for $FP$. The estimator $\widehat{\boldsymbol{\beta}}$ of $\boldsymbol{\beta}$ is obtained by solving the following weighted estimating equations for $\boldsymbol{\beta}$

$$\tilde{S}(\boldsymbol{\beta}) = \frac{\partial \tilde{l}(\boldsymbol{\beta})}{\partial \boldsymbol{\beta}} = \sum_{i \in s_c} (1 - p_i) \boldsymbol{x}_i - \sum_{i \in s_s} d_i p_i \boldsymbol{x}_i = \boldsymbol{0}.$$

According to Equation (4.1), the participation rate $\pi_i^{(c)}$ for unit $i \in s_c$ is estimated by $\hat{\pi}_i^{(c)} = \frac{\hat{p}_i}{1-\hat{p}_i}$, with $\hat{p}_i = \text{expit}(\widehat{\boldsymbol{\beta}}^T \boldsymbol{x}_i)$ being the estimated PS under the propensity model (4.2).

### 4.1 Inverse PS Weighting (IPSW) Method as a PS-Based Weighting Method

The IPSW method uses the inverse of estimated participation rate as the pseudo-weight, i.e., $w_i^{IPSW} = 1/\hat{\pi}_i^{(c)}$. The corresponding IPSW estimator of finite population mean, $\mu = \frac{1}{N} \sum_{i \in FP} y_i$, is

$$\hat{\mu}^{IPSW} = \frac{\sum_{i \in s_c} w_i^{IPSW} y_i}{\sum_{i \in s_c} w_i^{IPSW}}.$$

### 4.2 A Weak Exchangeability Assumption for PS-Based Matching Methods

We relax the SEA for the PS-based matching methods under the framework described above by using $p = P(R = 1 \mid \boldsymbol{x})$ (Rosenbaum and Rubin, 1983) or $q = \text{logit}(p)$ (Rubin and Thomas, 1992) as the matching score to measure the similarity among the cohort and survey units.

In general, the cohort $s_c$ is not representative of the finite population $FP$ because $s_c$ is not a random sample from $FP$. Therefore there are no sample weights for the cohort to use to equalize the distributions of covariates $x$ in $s_c$ and $FP$ if they differ. Thus, the pseudo-weights, created using the $x$, serve as the sample weights for $s_c$ to weight the $x$ distribution in $s_c$ up to that in $FP$.

The matching methods classify the cohort and survey individuals into "matching groups" with similar $x$-distributions (as measured by certain matching scores), and then distribute the survey weights (evenly by the PSAS method or fractionally by the KW method) to the matched cohort units. As a result, the marginal $x$-distribution in the pseudo-weighted $s_c$ becomes closer to the $x$-distribution in the $FP$ (estimated by the sample weighted $s_s$). The balancing score, defined below, should be used to group (or match) cohort and survey units so that the individuals sharing the same balancing score have the same $x$-distribution in $s_c$ and in $FP$. The balancing score $b(x)$ is a function of covariates $x$ such that the conditional distribution of $x$ given $b(x)$ is the same in the $s_c$ as that in the $FP$. We use the notation in Rosenbaum and Rubin (1983)

$$x \perp\!\!\!\perp R | b(x), \qquad (4.5)$$

where $R$ is defined in Equation (4.3). The coarsest balancing score is $p = \Pr(R = 1 | x)$, or any one-to-one functions of $p$, e.g., the participation rate $\pi^{(c)} = \frac{p}{1-p}$ or $q = \text{logit}\, p$ (Rubin and Thomas, 1992), which can be estimated from the propensity model (4.2) fitted to the combined ($s_c$ vs **weighted** $s_s$) sample.

For estimation of the finite population mean $\mu = \frac{1}{N}\sum_{i \in FP} y_i$, the requirement that the matching score should be a balancing score satisfying (4.5) can be relaxed to the WEA

$$E\{y | b(x), s_c\} = E\{y | b(x), FP\}, \qquad (4.6)$$

where $E$ is the expectation with respect to the distribution of $y$ in $FP$. The matching scores $p$ and $q$ satisfy Weak Exchangeability Assumption (WEA) (4.6) because they are balancing scores defined in (4.5). Since $p \in (0,1)$ and the distribution of $p$ is typically very right skewed as the participation rate of the cohort is small, then very small differences in $p$ may be resulted from large differences in covariates $x$, which can bias estimators when PS-based matching methods are applied. These boundary problems can be avoided by using $q = \text{logit}(p)$ (Rubin and Thomas, 1992) as the matching score.

## 4.3 Applying WEA to Kernel Weighting (KW) Approach

The kernel weighting (KW) approach, as a special case of PS-based matching methods, has been proved to provide consistent estimators of finite population means under SEA along with standard conditions (Wang et al., 2020), whereas other PS-matching methods such as PSAS may require more unrealistic conditions for consistent estimators. In this section, we propose an enhanced KW (referred to as KW.W) method by applying the WEA under the framework in Section 4.2, and provide statistical properties of KW.W estimators of finite population means.

Similar to the KW method, the KW.W method provides pseudo-weights, denoted by $w_i^{KW.W}$ for each individual $i \in s_c$, but the KW.W method uses $q = \text{logit}(p) = \boldsymbol{\beta}^T \boldsymbol{x}$ as a matching score, with $\boldsymbol{\beta}$ estimated under Model (4.2) fitted to the combined ($s_c$ vs. **weighted** $s_s$) sample by maximizing the pseudo-loglikelihood (4.4). Denote the estimated logit of propensity scores to be $q_i^{(c)}$ and $q_j^{(s)}$ for $i \in s_c$ and $j \in s_s$, respectively. The KW.W pseudo-weight, $w_i^{KW.W}$ for $i \in s_c$, is calculated as

$$w_i^{KW.W} = \sum_{j \in s_s} \left\{ \frac{K\left\{\left(q_i^{(c)} - q_j^{(s)}\right)/h\right\}}{\sum_{i \in s_c} K\left\{\left(q_i^{(c)} - q_j^{(s)}\right)/h\right\}} \cdot d_j \right\}, \qquad (4.7)$$

where $K(\cdot)$ is a zero-centered kernel function (Epanechnikov, 1969) (e.g. standard normal, or triangular density), $h$ is the bandwidth corresponding to the selected kernel function. The KW.W estimator of $\mu$ is $\hat{\mu}^{KW.W} = \frac{\sum_{i \in s_c} w_i^{KW.W} y_i}{\sum_{i \in s_c} w_i^{KW.W}}$.

**Theorem 1.** *Under the WEA* (4.6)*, conditions* **A1**, **A2** *and* **C1-C5** *in the Appendix A.1, the KW.W estimate of the finite population mean* $\hat{\mu}^{KW.W} = \mu + O_p\left(n_c^{-1/2}\right)$. *Assuming the logistic regression* *model* (4.2) *for the propensity scores* $p = Pr(R = 1 \mid \boldsymbol{x})$, *and under conditions* **C6**, **C8**, **C9** *in the Appendix A.1, we have the finite population variance* $Var(\hat{\mu}^{KW.W}) = V^{KW.W} + o(n_c^{-1})$, *where*

$$V^{KW.W} = N^{-2} \sum_{i \in FP} \pi_i^{(c)}\left(1 - \pi_i^{(c)}\right)\{w_i^{KW.W}(y_i - \mu) - (1 - p_i)\boldsymbol{b}^T\boldsymbol{x}_i\}^2 + \boldsymbol{b}^T D \boldsymbol{b}$$

*with* $\boldsymbol{b}^T = \left\{\sum_{i \in FP} \pi_i^{(c)}(y_i - \mu)\frac{\partial w_i^{KW.W}}{\partial \boldsymbol{\beta}}\right\}\{\sum_{i \in FP} p_i \boldsymbol{x}_i \boldsymbol{x}_i^T\}^{-1}$, $D = N^{-2} V_p\left(\sum_{i \in FP} \delta_i^{(s)} d_i p_i \boldsymbol{x}_i\right)$, *and* $V_p$ *denoting the design-based finite population variance under the probability sampling design for* $s_s$. *Notice that* $\frac{\partial w_i^{KW.W}}{\partial \boldsymbol{\beta}}$ *depends on the choice of kernel function* $K(\cdot)$ *(proof in Appendix A.2)*.

A consistent sample estimator of $V^{KW.W}$ can be obtained by substituting the finite population quantities by consistent sample estimators (Appendix B.1-B.2)

### 4.4 Satisfying SEA under WEA

Using matching score $\tilde{p}$ defined in (3.1) for matching methods may yield more efficient estimators, as the survey weights are not considered in the PS estimation, but it can bias estimates of population means if SEA is invalid. Under the propensity model (3.2), we have $\boldsymbol{x} \perp\!\!\!\perp T \mid \tilde{p}$ with $T$ indicating the group membership of $s_c$ versus $s_s$ (Rosenbaum and Rubin, 1983), and $E\{y \mid \tilde{p}, s_c\} = E\{y \mid \tilde{p}, s_s\}$. However, $\tilde{p}$ may not satisfy the second equality in the SEA, i.e., $E\{y \mid \tilde{p}, s_c\} \neq E\{y \mid \tilde{p}, FP\}$. For example, assuming the propensity score $p$ in Model (4.2) satisfies WEA (4.6). if $\tilde{p}$ is a one-to-one function (e.g. cases 1 and 2 in Figure 1) or many-to-one function (e.g. case 3 in Figure 1) of $p$, the second equality of the SEA $E\{y \mid \tilde{p}, s_c\} = E\{y \mid \tilde{p}, FP\}$ holds as $E\{y \mid p, s_c\} = E\{y \mid p, FP\}$. Otherwise $\tilde{p}$ will not satisfy the second equality the SEA (e.g. cases 4 and 5 in Figure 1). As a result, using matching scores of $\tilde{p}$ (or $\tilde{q} = \text{logit}\,\tilde{p}$) in the matching methods can produce biased estimators of population means.

## 4.5 Improving Efficiency of the IPSW and KW Estimators by Scaling the Survey Weights in Propensity Estimation

The IPSW and KW.W estimators of population means can be inefficient because of the generally large variability of the weights among the combined $s_c$ (with common weight of one) and sample weighted $s_s$ (with survey weight of $d_i$, $i \in s_s$) (Wang et al., 2020). Scaling weights has been suggested to improve efficiency of estimators in population-based case-control studies when the weights are highly variable among cases and controls (Scott and Wild, 1986; Li, Graubard, and DiGaetano, 2010). Following the rationale of Scott and Wild (1986), we propose scaling the survey weights $\{d_i, i \in s_s\}$ by the scaling factor $a = \frac{n_s}{\sum_{i \in s_s} d_i}$ and denote the scaled weight for the survey unit $i \in s_s$ by $d_i^* = a \cdot d_i$, so that $\sum_{i \in s_s} d_i^* = n_s$. The propensity model (4.2) is fitted to the combined ($s_c$ vs. **scaled-weighted** $s_s$) sample and the pseudo log-likelihood (4.4) with the $d_i$ replaced by $d_i^*$ is maximized to solve for $\boldsymbol{\beta}$. The resulting estimator is denoted by $\widehat{\boldsymbol{\beta}}^*$.

**Lemma 1.** $\widehat{\boldsymbol{\beta}}^*$ is a consistent estimator of $\boldsymbol{\beta}^* = \boldsymbol{\beta} + \log a \cdot \boldsymbol{e}_1$, where $\boldsymbol{e}_1 = (1, 0, \cdots, 0)^T$, $a$ is the scaling factor for survey weights, and $\boldsymbol{\beta}$ is the vector of regression coefficients defined in Model (4.2) (see proof of **Lemma 1** in the Appendix A.3).

Lemma 1 shows that rescaling survey weights in the combined sample for the propensity modeling only affects the intercept of the coefficients, which can be corrected by the offset of $\log a$. Therefore, as shown in **Theorem 2**, the resulting IPSW.S and KW.S estimators of the population mean $\mu$ when using the weights $d_i^*$ for propensity estimation, denoted by $\hat{\mu}^{IPSW.S}$ and $\hat{\mu}^{KW.S}$, are also consistent estimators of $\mu$ as were $\hat{\mu}^{IPSW}$ and $\hat{\mu}^{KW.W}$.

**Theorem 2** *Under the WEA* (4.6), *conditions* **A1**, **A2** *and* **C1-C5** *in the appendix, and assuming the logistic regression model* (4.2), *we have* $\hat{\mu}^* = \mu + O_p(n_c^{-1/2})$, *with* $\hat{\mu}^*$ *being either* $\hat{\mu}^{IPSW.S}$ *or* $\hat{\mu}^{KW.S}$. *Under conditions* **C7-C9** *in the Appendix, we have the finite population variance* $Var(\hat{\mu}^*) = V^* + o(n_c^{-1})$, *with*

$$V^* = N^{-2} \sum_{i \in FP} \pi_i^{(c)}\left(1 - \pi_i^{(c)}\right)\{\widetilde{w}_i^*(y_i - \mu) - (1 - p_i^*)\boldsymbol{b}^{*T}\boldsymbol{x}_i\}^2 + \boldsymbol{b}^{*T}D^*\boldsymbol{b}^*,$$

*where $V^*$ can be $V^{IPSW.S}$ or $V^{KW.S}$ depending on the choice of $\{\widetilde{w}_i^*, i \in FP\}$ being a set of IPSW.S or KW.S pseudo weights, $\boldsymbol{b}^*$ and $D^*$ are obtained by replacing $w_i^{KW.W}$, $p_i$ and $d_i$ with $\widetilde{w}_i^*$, $p_i^*$ and $d_i^*$ in $\boldsymbol{b}$ and $D$ defined in* **Theorem 1**, *respectively (proof and details in Appendix A.4).*

## 5 SIMULATIONS

### 5.1 Generating the Finite Population

We generated a finite population ($FP$) of size $N = 200,000$, with four covariates $x_1 \sim N(1,1)$, $x_2 \sim N(1,1)$, $x_3 \sim \text{LogNormal}(0, 0.7)$, and $x_4$ (=1 if $x_1 + x_2 > 2$; 0 otherwise). Note $x_1$ and $x_2$ are correlated with $x_3$, but independent of $x_4$. The outcome $y$ for $i \in FP$ were generated by $y_i = 2 + x_{1,i} + x_{2,i} + x_{4,i} + \epsilon_i, i \in FP$, where the error terms $\epsilon_i$ were independent and identically distributed (iid) as $N(0, 1)$. The finite population mean of $y$ is $\mu = 4.50$.

For each $i \in FP$, we created two variables, $x_1^*$ and $x_1^{**}$ as functions of $x_1$: $x_{1,i}^* = x_{1,i} + 0.15x_{1,i}^3$, and $x_{1,i}^{**}$ was defined as a categorical variable (=1 if $x_{1,i} \leq 10^{th}$ percentile; 2 if $10^{th} < x_{1,i} \leq 40^{th}$ percentiles; 3 if $40^{th} < x_{1,i} \leq 70^{th}$ percentiles; 4 if $70^{th} < x_{1,i} \leq 90^{th}$ percentiles; and 5 if $x_{1,i} > 90^{th}$ percentile of $x_1$ in the $FP$). The variables of $x_1^*$ and $x_1^{**}$ were used in the simulations as a substitute of the covariate $x_1$ to reflect cases when $x_1$ is not available but related variables are available.

### 5.2 Sampling from the Finite Population to Assemble the Survey Sample and Cohort

A cohort of size $n_c = 2,400$ individuals was randomly selected from the $FP$ by Probability Proportional to Size (PPS) sampling with the measure of size (MOS) for individual $i \in FP$ defined by $m_i^{(c)} = \exp(\alpha_1 x_{1,i} + \alpha_2 x_{2,i} + \alpha_3 x_{3,i})$, where $\boldsymbol{\alpha} = (\alpha_1, \alpha_2, \alpha_3) = (0.6, 0.15, 0.24)$. The sample weight (i.e., the reciprocal of the selection probability) for individual $i$ in the cohort was $w_i^{(c)} = \frac{\sum_{i=1}^N m_i^{(c)}}{n_c \cdot m_i^{(c)}}$. A survey sample of size $n_s = 2{,}000$ individuals was sampled independently of the sampling of the cohort where a similar PPS sampling design was used, but with a different MOS $m_i^{(s)} = \exp(\gamma_1 x_{1,i} + \gamma_2 x_{2,i} + \gamma_3 x_{3,i})$.

Under the PPS sampling described above, the true propensity model of a population unit included in $s_c$ vs. $FP$ (assumed by the IPSW and KW.W methods), and that in $s_c$ vs. $s_s$ (assumed by the original KW) were

$$\text{logit}\{P(i \in s_c | i \in s_c \cup^* FP)\} = \beta_0 + \boldsymbol{\beta}_1^T x_i, \tag{7.1}$$

and

$$\text{logit}\{P(i \in s_c | i \in s_c \cup^* s_s)\} = \tilde{\beta}_0 + \tilde{\boldsymbol{\beta}}_1^T x_i, \tag{7.2}$$

respectively, where $\boldsymbol{\beta}_1 = \boldsymbol{\alpha}$, $\tilde{\boldsymbol{\beta}}_1 = \boldsymbol{\alpha} - \boldsymbol{\gamma}$; $\beta_0$ and $\tilde{\beta}_0$ are the intercepts (Appendix A.5) and $x_i = (x_{1,i}, x_{2,i}, x_{3,i})^T$. The two propensity models have the same functional form so that the proposed PS-based weighting and matching methods can be fairly compared.

Values of $\boldsymbol{\gamma} = (\gamma_1, \gamma_2, \gamma_3)$ in MOS of survey sample selection can be varied to control for the validity of the SEA assumed by the original KW method (Wang et al. 2020). We considered two scenarios with $\boldsymbol{\gamma} = (-0.4, -0.1, 0.16)$ in Scenario 1, and $\boldsymbol{\gamma} = (-0.65, 0.2, 0)$ in Scenario 2.

Following Section 4.4, we made a scatter plot of $\tilde{q} = \tilde{\boldsymbol{\beta}}_1^T x$ estimated from the true propensity model (7.2) versus $q = \boldsymbol{\beta}_1^T x$ estimated from the true model (7.1) in both scenarios, where $x = (x_1, x_2)$. As shown in Figure 2, SEA is valid under model (7.2) in scenario 1, but invalid in scenario 2.

## 5.3 Evaluating Criteria

We examined performance of the five PS-based estimators of $\mu$: two IPSW estimates ($\hat{\mu}^{IPSW}$, $\hat{\mu}^{IPSW.S}$) and three KW methods ($\hat{\mu}^{KW}$, $\hat{\mu}^{KW.W}$, $\hat{\mu}^{KW.S}$), which are compared to the naïve unweighted cohort estimator ($\hat{\mu}^{Naive}$) and the weighted survey estimator ($\hat{\mu}^{SVY}$). We used criteria of relative bias (%RB), empirical variance ($V$), mean squared error (MSE) of the estimators, defined by

$$\text{\%RB} = \frac{1}{B}\sum_{b=1}^{B} \frac{\hat{\mu}^{(b)} - \mu}{\mu} \times 100, \quad V = \frac{1}{B-1}\sum_{b=1}^{B}\left\{\hat{\mu}^{(b)} - \frac{1}{B}\sum_{b=1}^{B}\hat{\mu}^{(b)}\right\}^2, \quad \text{MSE} = \frac{1}{B}\sum_{b=1}^{B}\left\{\hat{\mu}^{(b)} - \mu\right\}^2,$$

where $B = 10{,}000$ is the number of simulations, $\hat{\mu}^{(b)}$ is the estimate of $FP$ mean $\mu$, obtained from the $b$-th simulated samples.

For each mean estimator, we evaluated two variance estimators, i.e., the proposed Taylor linearization (TL) estimator (described in Appendix B.2) and the Jackknife replication (JK) estimator (Appendix B.3), using the variance ratio (VR), and coverage probabilities (CP) of the corresponding 95% confidence intervals, defined by

$$\text{VR} = \frac{\frac{1}{B}\sum_{b=1}^{B}\hat{v}^{(b)}}{V} \times 100, \quad \text{and} \quad \text{CP} = \frac{1}{B}\sum_{b=1}^{B} I\left(\mu \in CI^{(b)}\right),$$

where $\hat{v}^{(b)}$ is the variance estimate of $\hat{\mu}^{(b)}$, and $CI^{(b)} = \left(\hat{\mu}^{(b)} - 1.96\sqrt{\hat{v}^{(b)}},\ \hat{\mu}^{(b)} + 1.96\sqrt{\hat{v}^{(b)}}\right)$ is the 95% confidence interval from the $b$-th simulated sample.

## 5.4 Results under Scenario 1: the valid SEA

Table 1 shows the results under the SEA. The unweighted naïve cohort mean, $\hat{\mu}^{Naive}$, was biased by 20.97% while the survey estimate $\hat{\mu}^{SVY}$ was approximately unbiased. All KW and IPSW methods yielded approximately unbiased estimates of $\mu$. The KW estimator, $\hat{\mu}^{KW}$, had the smallest variance because no sample weights were considered in estimating PS's. The $\hat{\mu}^{KW}$ required the SEA, and naturally had the smallest MSE and maintained the nominal CP in Scenario 1.

## 5.5 Results under Scenario 2: the invalid SEA

Table 2 shows results under four fitted propensity models that includes different sets of covariates. Under the correctly specified propensity model (*Model T*), though the KW estimator, $\hat{\mu}^{KW}$, had the smallest variance among the five pseudo-weighted estimators, it had the largest bias, leading to low CP and the largest MSE, whereas $\hat{\mu}^{KW.W}$, and $\hat{\mu}^{KW.S}$ had smaller biases. Similar to the results in Table 1, scaling survey weights in the propensity model yielded more efficient estimates, especially for the IPSW method. Though $\hat{\mu}^{IPSW.S}$ had ~60% smaller variance than $\hat{\mu}^{IPSW}$, it was not as efficient as $\hat{\mu}^{KW.S}$. As a result, $\hat{\mu}^{KW.S}$ performed the best in terms of MSE.

Excluding $x_4$ in the propensity model (*Model U*) did not affect the extent of the bias of the estimates because $x_4$ was uncorrelated with the outcome variable $y$. However, the empirical variances of $\hat{\mu}^{IPSW}$ and $\hat{\mu}^{KW.W}$ were substantially reduced compared to the variances under *Model T* (similar findings as in Wang et al., 2020). In contrast, the variances of $\hat{\mu}^{IPSW.S}$ and $\hat{\mu}^{KW.S}$ were nearly unchanged.

In *Model $M_1$*, the true covariate $x_1$ in *Model U* was substituted by $x_1^*$. The IPSW estimators $\hat{\mu}^{IPSW}$ and $\hat{\mu}^{IPSW.S}$ were biased because the cohort participation rates cannot be accurately estimated from *Model $M_1$*. However, the matching methods with matching scores $\hat{q} = \hat{\beta}_1 x_1^* + \hat{\beta}_2 x_2$ still worked well because it was close to a one-to-one function of the true participation rate. As a result, the KW estimates were less biased than the IPSW estimates. Furthermore, by scaling survey weights, $\hat{\mu}^{KW.S}$ performed best with smallest bias, variance, and nearly nominal CP.

In contrast, *Model $M_2$* substituted $x_1$ by $x_1^{**}$, which was a categorical variable that was coarser than $x_1$ in *Model U*. This model did not accurately estimate the cohort participation rates or provide an adequate balancing score used for matching, because individuals in the same category of $x_1^{**}$ took on the same values of the matching scores $\hat{q}$ and were incorrectly assigned the same pseudo-weights. Hence, all of the pseudo-weighted estimates were biased.

The TL variance estimates were close to the truth (with VR ≈ 1) for all estimates except for $\hat{\mu}^{IPSW}$ with its VR<<1. This result is due to the finite sample bias caused by the large variability of the sample weights in the combined ($s_c$ vs. **weighted** $s_s$) sample with the common value of one for the cohort weights vs. values ranging from 23 to 618 for the survey weights (same findings in Li, Graubard, and DiGaetano 2010; Landsman and Graubard, 2012). In contrast, the TL variance estimate for $\hat{\mu}^{IPSW.S}$ worked well since the variability of the weights in the combined ($s_c$ vs. scaled-weighted $s_s$) sample was reduced. The scaled survey sample weights range from 0.2 to 5.1.

The JK method consistently had larger estimates of variances compared to the TL variance estimates (similar results were shown by Efron and Gong, 1983), and the JK estimates were more accurate for the variance of $\hat{\mu}^{IPSW}$. However, in some of the simulations the JK overestimated the variance of $\hat{\mu}^{KW}$ and $\hat{\mu}^{KW.S}$ due to highly variable covariates in the propensity model (*Model $M_1$*).

In summary, scaling the survey weights substantially not only decreased the variance of the mean estimates, but also reduced the finite sample bias of the TL variance estimates. The resulting estimates, $\hat{\mu}^{IPSW.S}$ and $\hat{\mu}^{KW.S}$ outperformed $\hat{\mu}^{IPSW}$ and $\hat{\mu}^{KW}$, respectively. The proposed $\hat{\mu}^{KW.S}$ generally had the smallest variance among the four methods, and its variance changed least among all the four estimates as the fitted propensity model varied. The proposed $\hat{\mu}^{KW.S}$ had the smallest MSE when the propensity model was appropriately specified (*Models T* and *U*). Under *Model $M_1$* when the variable(s) in the fitted propensity model was no coarser than the correct variable(s), $\hat{\mu}^{KW.S}$ was robust to model misspecification, and therefore unbiased and more efficient than $\hat{\mu}^{IPSW.S}$. Under *Model $M_2$*, the performance of $\hat{\mu}^{KW.S}$ and $\hat{\mu}^{IPSW.S}$ was comparable and $\hat{\mu}^{IPSW.S}$ had slightly smaller MSE than $\hat{\mu}^{KW.S}$ due to the smaller bias.

# 6 DATA ANALYSIS: The U.S. National Health and Nutrition Examination Survey

For our example, we used the Third U.S. National Health and Nutrition Examination Survey (NHANES III) as the volunteer-based "cohort" (ignoring sample weights) and the contemporaneous U.S. National Health Interview Survey (NHIS) as the reference survey. This example has several advantages for illuminating the performance of our methodology. The "cohort" and reference survey have approximately the same target population, data collection mode, and questionnaires. This ensures that when applying our methodology to the "cohort in an effective manner we should be able approximately recover US-representative estimates, enabling us to characterize the performance of our methodology with real data. Although problems with misaligned target populations and data harmonization are serious practical issues, they are beyond the scope of our methodology.

We estimated prospective 15-year all-cause mortality rates for adults in the US using the adult sample of household interview part of NHANES III conducted in 1988-1994, with sample size = 20,050. NHANES III is partly a cross-sectional household interview survey, and partly a medical examination survey of the civilian, non-institutionalized population of the United States. NHANES III oversampled poverty areas, children under age 5, adults aged 60 and over, non-Hispanic blacks, and Mexican Americans (Ezzati et al., 1992). The coefficient of variation (CV) of sample weights is 125%, indicating highly variable selection probabilities, and thus potential low representativeness of the unweighted sample. We ignored all complex design features of NHANES III and treated it as a cohort. For estimating mortality rates, we approximate that the entire sample of NHANES III was randomly selected in 1991 (the midpoint of the data collection time period).

For the reference survey, we used the 1994 NHIS respondents to the supplement for monitoring achievement of the Healthy People Year 2000 objectives, aged 18 and older (sample size = 19738). NHIS is also a cross-sectional household interview survey of the same target finite population as the NHANES III. The 1994 NHIS had a multistage stratified cluster sample design, with over sampling of the aged, low income, and Black and Hispanic populations (Massey et al., 1989). There were 125 strata and 248 pseudo-PSUs in the sample. We collapsed strata with only one PSU with the next nearest strata for variance estimation purpose (Hartley, Rao, and Kiefer, 1969). The CV of sample weights in 1994 NHIS sample is 58%. NHANES III and NHIS were linked to National Death Index (NDI) for mortality (National Center for Health Statistics, 2013), allowing us to quantify the relative bias of unweighted NHANES estimates, assuming the NHIS estimates as the gold standard.

We first compare the distributions of selected common covariates in the two samples (Table 3). As expected, the covariates in the weighted samples of NHANES and 1994 NHIS have very close distributions because both weighted samples represent approximately the same finite population. There are two exceptions: (1) education level, probably due to differences in how the question was asked in the two surveys; (2) health status, which was self-reported in NHANES but reported by the proxy of household representative in NHIS. As expected, the covariates distribute quite differently in the *unweighted* NHANES from the weighted samples, especially for design variables such as age, race/ethnicity, poverty, and region.

We use an AIC-based stepwise procedure (Lumley, 2020) to choose the propensity model fitted to combined sample of unweighted NHANES and *weighted* NHIS. This initially included main effects of common demographic characteristics (age, sex race/ethnicity, region, and marital status), socioeconomic status (education level, poverty, and household income), tobacco usage (smoking status, and chewing tobacco), health variables (body mass index [BMI], and self-reported health status), a quadratic term for age, and all two-way interactions. Table C.2 in Appendix C shows the final propensity models fitted to the weighted sample (for IPSW and KW.W), scaled weighted sample (for IPSW.S and KW.S) and unweighted sample (for KW).

To evaluate the performance of the five PS-based methods, we used relative difference from the NHIS estimate %RD$=\frac{\hat{\mu}-\hat{\mu}^{NHIS}}{\hat{\mu}^{NHIS}} \times 100$, bias reduction from the naïve (unweighted) NHANES estimates %BR$=\frac{\hat{\mu}^{Naive}-\hat{\mu}}{\hat{\mu}^{Naive}-\hat{\mu}^{NHIS}} \times 100$, TL variance estimate ($V$), and estimated MSE $= (\hat{\mu} - \hat{\mu}^{NHIS})^2 + V$, which treated the NHIS estimates as truth.

Table 5 shows that the weighted 1994 NHIS and the sample-**weighted** NHANES III estimates (TW) of 15-year all-cause mortality were very close (%RD = 2.6% for overall estimate, and %RD = 4.5-7.3% on average for the estimates by subgroups). In contrast, the naïve NHANES III estimate of overall mortality was ~52% biased from the NHIS estimate because older people who have higher mortalities were oversampled, and the bias in subgroup-specific mortality reached 96.8% for Non-Hispanic Whites. All KW and IPSW methods substantially reduced the bias from the naïve estimates. The four methods that fit propensity models to the (scaled-) weighted sample (IPSW, IPSW.S, KW.W, and KW.S) provided the closest estimates. The bias in the naïve estimate of overall mortality was almost eliminated by the KW.W and KW.S methods (~96% bias removed). KW.S on average had the least bias for the subgroup-specific mortality among the four methods. Similar to the simulation results, KW.W and KW.S estimates had smaller variance (estimated by TL method) than the IPSW and IPSW.S estimates. As a result, the KW.S estimates had on average the smallest MSE.

Importantly, the original KW method had the largest bias in overall mortality (BR%=66% vs. ≥92.7%), but had least bias for age-specific mortality (BR%=43.2% vs. ≤31.0%) and achieved smallest MSE for most age groups. This paradox is can be explained by the invalidity of the SEA for overall mortality, but not for age-specific mortality. As shown in Table 4, the small biases in the KW estimates of age-specific mortality imply that the SEA held with $E(y \mid \text{age}, \tilde{p}, s_c)$ = $E(y \mid \text{age}, \tilde{p}, FP)$. As shown in Table 3, the KW pseudo-weighted age distribution in $s_c$ (unweighted NHANES sample) differed from that in $FP$ (represented by the weighted NHIS), indicating $P(\text{age} \mid \tilde{p}, s_c) \neq P(\text{age} \mid \tilde{p}, FP)$. As a result, the SEA was invalid for the overall mortality estimation using the original KW method, that is

$$E(y \mid \tilde{p}, s_c) = \sum_{\text{age}} \{E(y \mid \text{age}, \tilde{p}, s_c) P(\text{age} \mid \tilde{p}, s_c)\}$$

$$\neq \sum_{\text{age}} E(y \mid \text{age}, \tilde{p}, FP) P(\text{age} \mid \tilde{p}, FP) = E(y \mid \tilde{p}, FP).$$

This result is consistent with the findings in the simulations: the original KW estimates can have the smallest (or largest) MSE when the SEA is valid (or invalid). The best methods for overall mortality were KW.W and KW.S, both of which require only the WEA to hold.

The other four methods (IPSW, IPSW.S, KW.W, and KW.S) had similar estimated mortality rates. The IPSW estimates had the largest variances, followed by the IPSW.S estimates. The KW.S estimates had the smallest variances and MSE in most cases. The results of the JK and the TL variance estimates were similar in this example (results not shown).

## 7 DISCUSSION

We established a unified framework for both PS-based weighting and matching methods that improves estimates of finite population means from non-representative cohort data by using a reference representative survey sample of the target population. This unifying framework allows us to make three contributions. First, we identified the underlying Strong Exchangeability Assumption (SEA), implicitly assumed by existing PS-based matching methods, whose failure invalidates inference even if the PS-model is correctly specified. Our simulations and data example demonstrate that PS-based matching methods that rely on the SEA, such as the original KW estimator (Wang et al., 2020), have smallest MSE when the SEA holds, but have large bias when the SEA fails. Second, as a remedy, we proposed PS-based methods that require only a Weak Exchangeability Assumption (WEA). Third, we further improved the efficiency of PS-based estimates by scaling the survey weights to sum to the survey sample size. Scaling reduces the variance of the estimated PS's and thus markedly improves efficiency of the pseudo-weighted estimates, especially for the IPSW method. Our recommended method, kernel-weighting with scaling the survey weights (KW.S), is more robust by only requiring the WEA, yet the scaling ensures that it has smallest MSE.

For the variance estimation, we recommend the JK method for the IPSW estimates because our empirical results indicate that the TL method can have greater finite sample bias due to highly variable weights in the combined sample. However, both the JK and the TL methods provided good variance estimation for the IPSW.S estimates. The TL method is recommended for the KW.W and the KW.S estimates because the JK method can overestimate the variance.

In our data example, though the original KW method reduced most bias for age-specific mortality rate (~43% bias reduction v.s. 23%~31% for the other four methods), it had the largest bias for the overall mortality rate (~18% relative bias v.s. -3.8%~-2% for the other four methods). This is because SEA approximately holds for age-specific mortality, but failed for overall mortality (where WEA held). Among all the five PS-based methods, KW.S reduced most bias and obtained smallest MSE on average, due to robustness from matching, WEA and reduction of weight variability from kernel smoothing and scaling survey weights.

Our unifying framework codifies two other key assumptions generally taken for granted. Assumption **A1** ensures non-informative sampling of the cohort, allowing for correct estimation of participation rates. Assumption **A2** ensures that the cohort and the survey samples cover the same target finite population. Assumption **A1** is often reasonable, especially when the outcome is measured after the cohort is assembled, but assumption **A2** is generally violated, to some extent, in real data. For example, most cohort studies only recruit people in a few study centers in a target population (e.g., the US), while many surveys are representative of the target population. One solution is to use subgroups of the survey sample that are covered by the cohort as the reference so that the weighted cohort only represents a defined subpopulation. This problem of misaligned coverage between cohort and survey is a critical issue for future research.

There is much room for future research. More research is needed on propensity model selection and diagnostics. First, doubly robust estimators (Chen, Li, and Wu, 2019) are a promising approach to enhance robustness to propensity model misspecification by imputing the completely missing outcome into the survey based on the cohort. Second, diagnostics are need to assess if the SEA holds in an analysis. Finally, more work is needed to determine optimal scaling factors for rescaling the survey weights used in PS estimation that minimize the variance of the pseudo-weighted estimators.

**REFERENCE**

Brick, J. M. (2015). Compositional model inference. *JSM Proceedings (Survey Research Methods Section)*, 299-307.

Chen, Y., Li, P., and Wu, C. (2019). Doubly Robust Inference with Nonprobability Survey Samples. *Journal of the American Statistical Association*, 1-11.

Collins, R. (2012). What makes UK Biobank special? *The Lancet* **379**(9822), 1173-1174.

Duncan, G. J. (2008). When to promote, and when to avoid, a population perspective. *Demography*, 45(**4**), 763-784.

Efron, B., and Gong, G. (1983). A leisurely look at the bootstrap, the jackknife, and cross-validation. *The American Statistician*, **37**(1), 36-48.

Elliott, M. R., and Valliant, R. (2017). Inference for nonprobability samples. *Statistical Science*, **32**(2), 249-264.

Epanechnikov, V. A. (1969). Non-parametric estimation of a multivariate probability density. *Theory of Probability & Its Applications*, 14(1), 153-158.

Ezzati, T. M., Massey, J. T., Waksberg, J., Chu, A., and Maurer, K. R. (1992). Sample design: Third National Health and Nutrition Examination Survey. *Vital and health statistics.* **2** (113), 1-35.

Hartley, H. O., Rao, J. N. K., and Kiefer, G. (1969). Variance estimation with one unit per stratum. Journal of the American Statistical Association, 64(327), 841-851.

Landsman, V., and Graubard, B. I. (2013). Efficient analysis of case‐control studies with sample weights. *Statistics in Medicine*, **32**(2), 347-360.

LaVange, L. M., Koch, G. G., and Schwartz, T. A. (2001). Applying sample survey methods to clinical trials data. *Statistics in Medicine*, **20**(17-18), 2609-2623.

Lee, S. and Valliant, R. (2009). Estimation for volunteer panel web surveys using propensity score adjustment and calibration adjustment. *Sociological Methods & Research*, **37**:319–343.

Li, Y., Graubard, B., and DiGaetano, R. (2010), Weighting methods for population-based case-control study. *Journal of Royal Statistical Society C*, 60, 165–185.

Lumley, T., 2020. Package 'Survey.' (http://cran.r-project.org/web/packages/survey/survey.pdf).

Massey JT, Moore TF, Parsons VL, and Tadros W. (1989) Design and estimation for the National Health Interview Survey, 1985-94. National Center for Health Statistics. *Vital and health statistics* **2**(110).

National Center for Health Statistics. (2013). National Death Index user's guide. Hyattsville, MD: US Department of Health and Human Services, Centers for Disease Control and Prevention. (Available at: https://www.cdc.gov/nchs/data/ndi/ndi_users_guide.pdf)

Rivers, D. (2007, August). Sampling for web surveys. In *Joint Statistical Meetings*.

Rosenbaum, P. R., and Rubin, D. B. (1983). The central role of the propensity score in observational studies for causal effects. Biometrika, 70(1), 41-55.

Rubin, D. B., and Thomas, N. (1992). Characterizing the effect of matching using linear propensity score methods with normal distributions. *Biometrika*, **79**(4), 797-809.

Scott, A. J., and Wild, C. J. (1986). Fitting logistic models under case-control or choice based sampling. Journal of the Royal Statistical Society: Series B (Methodological), 48(2), 170-182.

Valliant, R., and Dever, J. A. (2011). Estimating propensity adjustments for volunteer web surveys. *Sociological Methods & Research* **40**(1), 105-137.

Wang, L., Graubard, B. I., Katki, H. A., and Li, Y. (2020). Improving external validity of epidemiologic cohort analyses: a kernel weighting approach. *Journal of the Royal Statistical Society Series A*, **183**(3), 1293-1311.

Table 1 Results from 10,000 simulated cohorts and survey samples under SEA.

| Estimator | %RB | V($\times 10^3$) | VR(TL) | VR(JK) | CP(TL) | CP(JK) | MSE($\times 10^3$) |
|---|---|---|---|---|---|---|---|
| $\hat{\mu}^{Naive}$ | 20.97 | 1.71 | 1.02 | | 0.00 | | 889.76 |
| $\hat{\mu}^{SVY}$ | 0.07 | 2.82 | 1.01 | 1.02 | 0.96 | 0.96 | 2.83 |
| **Model T** (True) logit{Pr($x$)} ~ $x_1, x_2, x_4$ | | | | | | | |
| $\hat{\mu}^{IPSW}$ | -0.12 | 9.96 | 0.92 | 1.01 | 0.94 | 0.95 | 9.99 |
| $\hat{\mu}^{IPSW.S}$ | 0.07 | 4.59 | 0.99 | 0.99 | 0.95 | 0.95 | 4.60 |
| $\hat{\mu}^{KW}$ | 0.18 | 2.54 | 1.07 | 1.06 | 0.96 | 0.95 | 2.61 |
| $\hat{\mu}^{KW.W}$ | 0.66 | 4.02 | 1.01 | 1.07 | 0.93 | 0.94 | 4.90 |
| $\hat{\mu}^{KW.S}$ | 0.63 | 3.12 | 1.03 | 1.07 | 0.93 | 0.93 | 3.92 |

Table 2 Results from 10,000 simulated cohorts and survey samples with each cohort and survey sample fitted to the correct propensity model and three misspecified propensity models with violated SEA.

| Estimator | %RB | V($\times 10^3$) | VR(TL) | VR(JK) | CP(TL) | CP(JK) | MSE($\times 10^3$) |
|---|---|---|---|---|---|---|---|
| $\hat{\mu}^{Naive}$ | 20.97 | 1.72 | 1.02 | | 0.00 | | 889.89 |
| $\hat{\mu}^{SVY}$ | 0.04 | 3.61 | 1.02 | 1.02 | 0.95 | 0.95 | 3.62 |
| **Model T** (True) logit{Pr($x$)} ~ $x_1, x_2, x_4$ | | | | | | | |
| $\hat{\mu}^{IPSW}$ | -0.36 | 14.73 | 0.86 | 1.03 | 0.93 | 0.94 | 14.99 |
| $\hat{\mu}^{IPSW.S}$ | 0.03 | 5.92 | 0.98 | 1.00 | 0.95 | 0.95 | 5.92 |
| $\hat{\mu}^{KW}$ | 4.84 | 2.66 | 0.92 | 1.04 | 0.01 | 0.02 | 50.03 |
| $\hat{\mu}^{KW.W}$ | 0.84 | 4.83 | 1.04 | 1.09 | 0.93 | 0.94 | 6.24 |
| $\hat{\mu}^{KW.S}$ | 0.65 | 3.55 | 1.02 | 1.08 | 0.93 | 0.93 | **4.39** |
| **Model U** (Underfitted) logit{Pr($x$)} ~ $x_1, x_2$ | | | | | | | |
| $\hat{\mu}^{IPSW}$ | -0.24 | 13.62 | 0.89 | 1.02 | 0.93 | 0.94 | 13.73 |
| $\hat{\mu}^{IPSW.S}$ | 0.04 | 5.76 | 0.98 | 1.00 | 0.95 | 0.95 | 5.77 |
| $\hat{\mu}^{KW.W}$ | 0.80 | 3.92 | 1.12 | 1.13 | 0.94 | 0.94 | 5.22 |
| $\hat{\mu}^{KW.S}$ | 0.57 | 3.39 | 1.02 | 1.08 | 0.93 | 0.94 | **4.05** |
| **Model $M_1$** (Mis-specified variable) logit{Pr($x$)} ~ $x_1^*, x_2$ | | | | | | | |
| $\hat{\mu}^{IPSW}$ | 4.91 | 16.94 | 0.62 | 1.08 | 0.44 | 0.55 | 65.61 |
| $\hat{\mu}^{IPSW.S}$ | 3.22 | 7.10 | 0.95 | 1.00 | 0.58 | 0.58 | 28.04 |
| $\hat{\mu}^{KW.W}$ | 0.58 | 4.54 | 1.08 | 1.45 | 0.87 | 0.95 | 5.22 |
| $\hat{\mu}^{KW.S}$ | 0.52 | 3.18 | 0.94 | 1.29 | 0.92 | 0.96 | **3.72** |
| **Model $M_2$** (Mis-specified variable) logit{Pr($x$)} ~ $x_1^{**}, x_2$ | | | | | | | |
| $\hat{\mu}^{IPSW}$ | 1.54 | 7.46 | 0.98 | 1.00 | 0.86 | 0.86 | 12.28 |
| $\hat{\mu}^{IPSW.S}$ | 1.58 | 4.58 | 0.99 | 1.00 | 0.82 | 0.82 | **9.62** |
| $\hat{\mu}^{KW.W}$ | 2.08 | 4.10 | 0.99 | 1.12 | 0.70 | 0.75 | 12.87 |
| $\hat{\mu}^{KW.S}$ | 2.04 | 3.62 | 0.92 | 1.10 | 0.65 | 0.72 | 12.01 |

Table 3 Relative difference of age group proportion estimates from the 1994 NIHS estimates

| Age Group | IPSW | IPSW.S | KW | KW.W | KW.S |
|---|---|---|---|---|---|
| 18-24 yrs | -8.2 | -8.3 | **-23.6** | -7.3 | -8.3 |
| 25-44 yrs | 4.5 | 3.1 | -4.3 | 2.9 | 2.5 |
| 45-64 yrs | -1.6 | 0.5 | 4.6 | -0.7 | 0.6 |
| 65-69 yrs | -11.7 | -9.0 | 5.4 | -10.6 | -8.8 |
| 70-74 yrs | -4.0 | -2.1 | **18.8** | -1.8 | 0.1 |
| >=75 yrs | 4.6 | 2.2 | **39.8** | 7.4 | 4.2 |
| Average | 5.8 | 4.2 | **16.1** | 5.1 | 4.1 |

Table 4 Estimates of all-cause 15-year mortality (overall, and by subgroups) with estimated variance and mean squared error

| | Est | %Relative Difference from the NHIS Estimate | | | | | | | TL Variance Estimate (× $10^5$) | | | | | MSE (× $10^5$) | | | | |
|---|---|---|---|---|---|---|---|---|---|---|---|---|---|---|---|---|---|---|
| | NHIS | TW | Naïve | KW | IPSW | IPSW.S | KW.W | KW.S | KW | IPSW | IPSW.S | KW.W | KW.S | KW | IPSW | IPSW.S | KW.W | KW.S |
| **Overall** | 17.6 | -2.6 | 52.2 | 17.7 | -3.8 | -3.2 | **-2.0** | -2.2 | 1.2 | 1.4 | 1.0 | 1.0 | **1.0** | 98.2 | 5.8 | 4.3 | **2.3** | 2.5 |
| (%BR) | | | | (66.0) | (92.7) | (93.8) | **(96.1)** | (95.8) | | | | | | | | | | |
| **Age group** | | | | | | | | | | | | | | | | | | |
| 18-24 yrs | 2.2 | -16.1 | **0.5** | -35.9 | -33.4 | -30.9 | -32.4 | **-30.2** | **0.8** | 0.8 | 0.9 | 0.8 | 0.9 | 7.2 | 6.4 | 5.7 | 6.1 | **5.5** |
| 25-44 yrs | 3.9 | -7.9 | 30.9 | **-4.5** | -14.8 | -14.3 | **-12.5** | -14.0 | **0.8** | 0.7 | 0.7 | 0.7 | **0.7** | **1.1** | 4.0 | 3.8 | **3.1** | 3.6 |
| 45-64 yrs | 17.7 | 5.8 | 30.6 | **1.3** | -3.8 | **-3.2** | -4.1 | -3.7 | 5.2 | 5.4 | 5.1 | **4.9** | 4.9 | **5.7** | 9.8 | **8.2** | 10.2 | 9.2 |
| 65-69 yrs | 45.5 | 0.9 | 9.5 | **-1.3** | -6.4 | -5.4 | -5.8 | **-4.7** | 35.9 | 35.6 | 34.4 | 33.5 | **33.3** | **39.5** | 120.8 | 95.1 | 103.3 | **78.1** |
| 70-74 yrs | 60.0 | 3.5 | 6.4 | **-0.4** | -1.4 | -1.3 | -1.3 | **-1.1** | 32.1 | 31.6 | 30.3 | 30.4 | **29.6** | **32.7** | 39.2 | 36.9 | 36.9 | 33.9 |
| >=75 yrs | 86.2 | 1.1 | 4.3 | 3.3 | 3.3 | 3.2 | 3.2 | **3.1** | 6.0 | 6.1 | 5.8 | 5.9 | **5.8** | 85.5 | 85.7 | 80.2 | 83.4 | **76.2** |
| Average | | 5.9 | 13.7 | **7.8** | 10.5 | 9.7 | 9.9 | 9.4 | 13.5 | 13.4 | 12.9 | 12.7 | **12.5** | 28.6 | 44.3 | 38.3 | 40.5 | **34.4** |
| (%BR) | | | | **(43.2)** | (23.3) | (29.2) | (27.7) | **(31.0)** | | | | | | | | | | |
| **Sex** | | | | | | | | | | | | | | | | | | |
| Male | 18.8 | -7.1 | 58.1 | 15.9 | **0.0** | 0.8 | 1.8 | 2.0 | 2.9 | 2.8 | 2.5 | **2.4** | 2.4 | 91.7 | 2.8 | **2.7** | 3.5 | 3.8 |
| Female | 16.5 | 1.9 | 46.5 | 21.0 | -6.5 | -6.8 | **-4.8** | -5.9 | 2.1 | 2.2 | 1.7 | 1.7 | **1.7** | 121.7 | 13.7 | 14.3 | **7.9** | 11.1 |
| Average | | 4.5 | 52.3 | 18.4 | **3.3** | 3.8 | 3.3 | 3.9 | 2.5 | 2.5 | 2.1 | 2.1 | **2.0** | 106.7 | 8.3 | 8.5 | **5.7** | 7.4 |
| (%BR) | | | | (64.8) | **(93.8)** | (92.7) | (93.7) | (92.5) | | | | | | | | | | |
| **Race** | | | | | | | | | | | | | | | | | | |
| NH-White | 18.7 | -1.7 | 96.8 | 17.8 | -1.9 | -1.8 | **-0.2** | -0.8 | 1.9 | 2.2 | 1.7 | 1.6 | **1.6** | 112.1 | 3.5 | 2.8 | **1.6** | 1.8 |
| NH-Black | 18.9 | -5.6 | 19.4 | 17.3 | -4.4 | -6.9 | **-4.0** | -7.0 | 3.8 | 6.6 | 5.7 | 3.0 | **2.7** | 110.9 | 13.6 | 22.9 | **8.6** | 20.3 |
| Hispanic | 10.2 | -9.1 | 62.2 | 13.2 | -15.3 | -9.3 | -13.0 | **-7.9** | 2.6 | 6.1 | 5.7 | **1.8** | 1.9 | 20.7 | 30.5 | 14.5 | 19.4 | **8.3** |
| NH-Other | 9.0 | -12.8 | 63.8 | **-14.1** | -31.1 | -23.8 | -24.2 | -19.0 | 19.2 | 14.6 | 14.4 | **13.2** | 14.4 | **35.2** | 92.1 | 59.9 | 60.2 | 43.5 |
| Average | | 7.3 | 60.5 | 15.6 | 13.2 | 10.5 | 10.4 | **8.7** | 6.9 | 7.4 | 6.9 | **4.9** | 5.1 | 69.7 | 34.9 | 25.1 | 22.5 | **18.5** |
| (%BR) | | | | (74.2) | (78.2) | (82.7) | (82.9) | **(85.7)** | | | | | | | | | | |

Figure 1 Scatter plots of linear propensity scores for assessing validity of SEA .

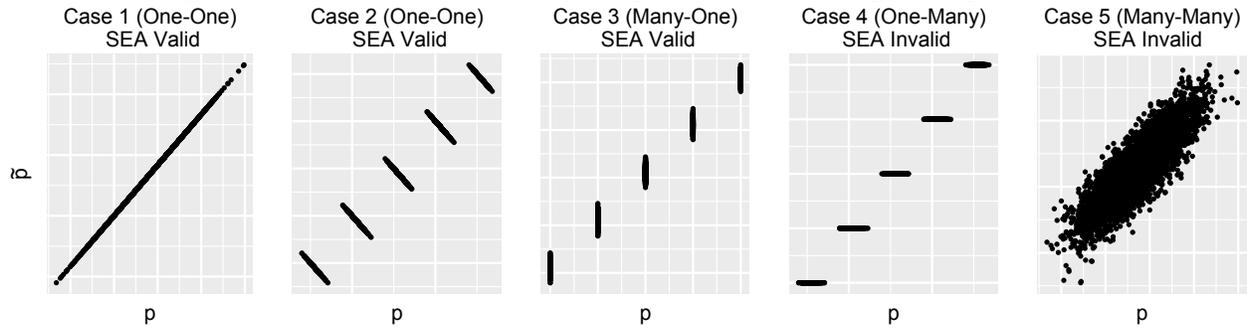

Figure 2 Scatter plots of linear propensity scores for assessing validity of SEA in two scenarios of simulations.

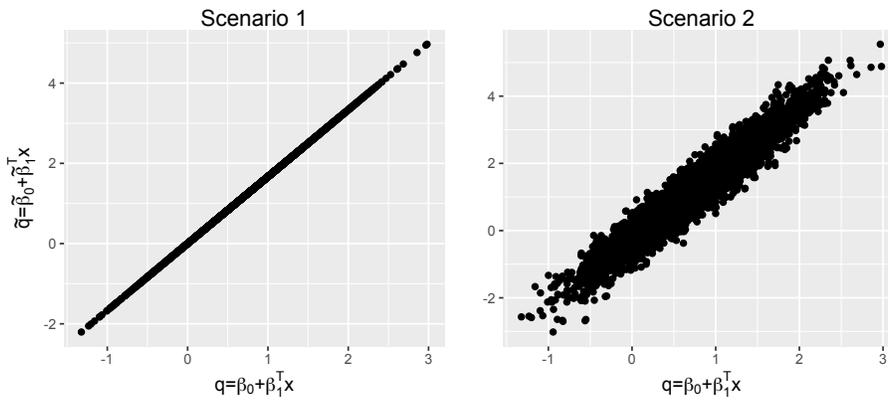